\documentclass[12pt]{article}
\usepackage{docstyle}
\usepackage{leftidx}

\renewcommand{\doctitle}{Assessing Wind Impact on Semi-Autonomous Drone Landings for In-Contact Power Line Inspection}

\begin{document}
    \title{\doctitle}
    \pagenumbering{roman}
    
    \begin{titlepage}
\begin{center}


\HRule\\[0.4cm]
{\huge\bfseries\doctitle}
\HRule\\[1.5cm]

\begin{tabulary}{\textwidth}{L L L C}
  \large\emph{Autors} & 
  \large\emph{Email} &
  \large\emph{Affiliation} \\
  Étienne Gendron & \href{mailto:etienne.gendron@usherbrooke.ca}{etienne.gendron@usherbrooke.ca} & University of Sherbrooke \\
  Marc-Antoine Leclerc & \href{mailto:mantoine.leclerc@usherbrooke.ca}{marc-antoine.leclerc@usherbrooke.ca} & University of Sherbrooke \\
  Samuel Hovington & \href{mailto:samuel.hovington@usherbrooke.ca}{samuel.hovington@usherbrooke.ca} & University of Sherbrooke \\
  Étienne Perron & \href{mailto:etienne.perron@cae.com}{etienne.perron@cae.com} & University of Sherbrooke \\
  David Rancourt & \href{mailto:david.rancourt2@usherbrooke.ca}{david.rancourt2@usherbrooke.ca} & University of Sherbrooke \\
  Alexis Lussier-Desbiens & \href{mailto:alexis.lussier.desbiens@usherbrooke.ca}{alexis.lussier.desbiens@usherbrooke.ca} & University of Sherbrooke\\
  Philippe Hamelin & \href{mailto:hamelin.philippe@hydroquebec.com}{hamelin.philippe@hydroquebec.com} & Hydro-Québec\\
  Alexandre Girard & \href{mailto:Alexandre.Girard2@USherbrooke.ca}{Alexandre.Girard2@USherbrooke.ca} & University of Sherbrooke\\
\end{tabulary}

\vfill

Corresponding author :\\
Étienne Gendron, etienne.gendron@usherbrooke.ca

\vfill

{\large \today}

\end{center}
\end{titlepage}
    \newpage
    
    \tableofcontents
    
    \pagenumbering{arabic}
    \newpage
    \textbf{Abstract}: In recent years, the use of inspection drones has become increasingly popular for high-voltage electric cable inspections due to their efficiency, cost-effectiveness, and ability to access hard-to-reach areas. However, safely landing drones on power lines, especially under windy conditions, remains a significant challenge. This study introduces a semi-autonomous control scheme for landing on an electrical line with the NADILE drone (an experimental drone based on original LineDrone key features for inspection of power lines) and assesses the operating envelope under various wind conditions. A Monte Carlo method is employed to analyze the success probability of landing given initial drone states. The performance of the system is evaluated for two landing strategies, variously controllers parameters and four level of wind intensities. The results show that a two-stage landing strategies offers higher probabilities of landing success and give insight regarding the best controller parameters and the maximum wind level for which the system is robust. Lastly, an experimental demonstration of the system landing autonomously on a power line is presented. \\


\section{Introduction}

The use of inspection drones has gained popularity in recent years due to their efficiency, cost-effectiveness, and ability to access hard-to-reach areas. Drone inspection technologies have revolutionized power line inspections by enhancing efficiency, safety, and accuracy while reducing costs. Drones equipped with advanced sensors and imaging systems remotely monitor power lines, capturing detailed visuals and detecting issues such as wear, corrosion, or damage. Light Detection and Ranging (LiDAR) sensors provide accurate 3D representations \parencite{Lee_2019}, enabling the detection of small-scale deformations and precise measurements. Additional technologies, like magnetoresistive sensors \parencite{Wu_2019} and hyperspectral imaging \parencite{Qingsheng_2021}, help identify hidden defects. A recent development involves using drones to deploy contact probes on energized power lines. The LineOhm \parencite{lavoie_resistance-measuring_2021} measures sleeve resistance, and the LineCore \parencite{Bellemare_2021} measures conductor corrosion, serving as two successful examples of contact probes deployed by drones. 
Deploying contact probes on power lines using drones presents multiple challenges. These include ensuring accurate navigation and landing in varying wind conditions, managing interference from power line electromagnetic fields, maintaining safety from high voltage electricity, and operating the drones manually. This last point is difficult to achieve at all times because of the height of the power lines, the terrain and the parallax effect of the cables. Also for this moment the pilot's input is crucial for ensuring the safety of the operation. Therefore, this study presents a semi-autonomous control scheme for landing a drone on electrical lines and evaluates the operating envelope under different wind conditions for a power line-adapted drone, maintaining the pilot's involvement in the process. 

\begin{figure}[h]
    \centering
    \includegraphics[width=0.8\textwidth]{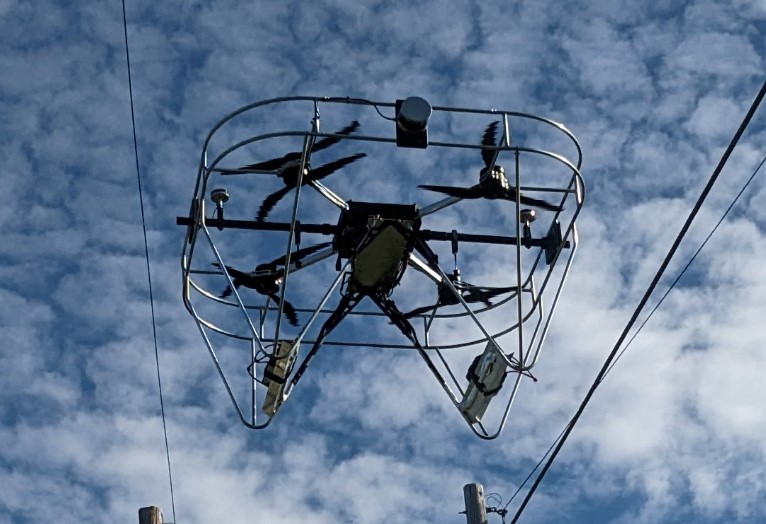}
    \caption{Drone NADILE in flight}
    \label{fig:LineDroneNadile}
\end{figure}

In an attempt to address these challenges, a drone prototype, named LineDrone, has been developed specifically for high voltage power line applications. The drone is designed to land and deploy contact probes directly onto the power lines. This innovative solution, developed by Hydro-Québec's robotics teams, facilitates efficient inspection of energized power lines while reducing accident risk and minimizing potential infrastructure damage. The drone prototype, designed for Autonomous Navigation of Drones and Interventions on Power Lines (acronym in French - NADILE), as shown in Figure \ref{fig:LineDroneNadile}, serves as the primary test-bed and evaluation model for this study.

\begin{figure}[ht]
    \centering
    \includegraphics[width=1\textwidth]{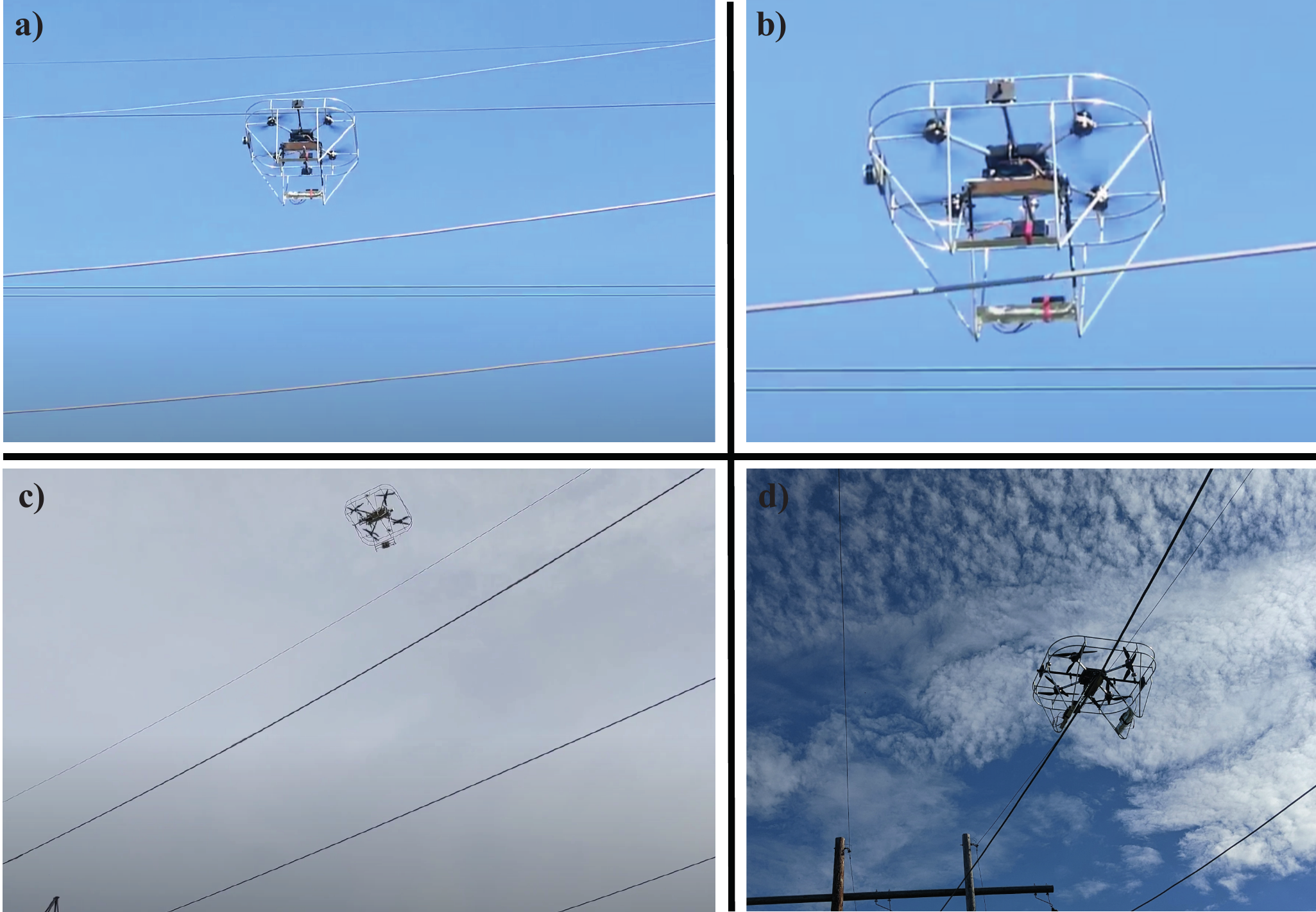}
    \caption{NADILE during landing on un-energized power lines mockup, in a), b) and c) it's difficult to say if the drone is aligned with the cable. d) shows clearly the drone on cable.} 
    \label{fig:LineDroneinaction}
\end{figure}

\begin{figure}[ht]
    \centering
    \includegraphics[width=1\textwidth]{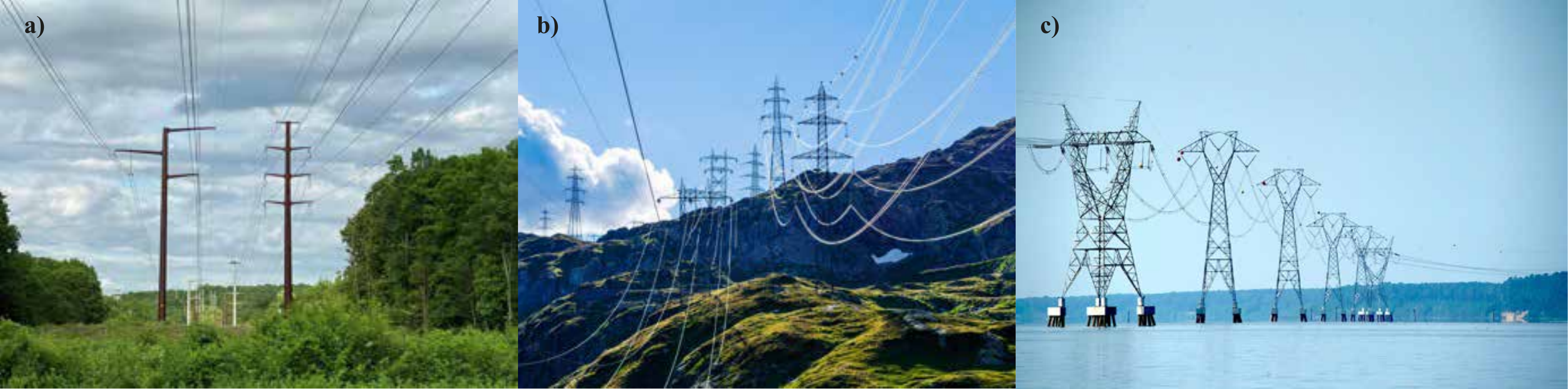}
    \caption{Difficult operating areas such, a) shrubs below the lines\parencite{turmelle_report_2022}, b) mountainous terrain \parencite{allain_lets_2017} and c) power-lines river crossing \parencite{npca_experts_2020}.}
    \label{fig:Diff_aera}
\end{figure}

Manual drone operation for this application poses challenges in maintaining clear line of sight for accurate positioning during landing, due to parallax effects and altitude as show in Figure \ref{fig:LineDroneinaction}. The current need to operate the drone from directly beneath the power line complicates its use in hard-to-reach areas, as illustrated in Figure \ref{fig:Diff_aera}.


This paper presents a semi-autonomous control scheme for landing a drone on an electrical line and evaluates its landing envelope. A methodology is developed to generate a map based on the drone's current state and wind conditions, enabling the calculation of landing envelopes, i.e., zones of states where initiating a landing has a high chance of success. These computations are utilized to assess two landing strategies under varying wind conditions, which provide insight into the drone's ability to land successfully. Additionally, the same calculations help establish transition rules for the autonomous landing state-machine, determining when to initiate or withhold the next landing phase. Lastly, the semi-autonomous control scheme is experimentally tested using the NADILE, showcasing secure semi-autonomous landings.

Section \ref{sec:StateofArt} explores existing landing strategies and analysis methods like Monte Carlo simulations that have inspired this study. Section \ref{sec:Methodology} details the methodology adopted for the landing strategies, control models, state-machines, and the drone model used in the simulations. Section \ref{sec: Results} presents the evaluation results of the NADILE's landing envelope, comparing different landing strategies under wind conditions and assessing various alignment controller gains. Section \ref{sec: fieldexp} concentrates on field tests of the semi-autonomous approach.

\section{States of the art} \label{sec:StateofArt}


Landing drones on cables for inspection has garnered significant attention in recent years. The paper \parencite{Miralles_2018} introduced a multicopter drone developed by the Hydro-Quebec (HQ) robotics team for landing on power lines and conducting inspections. The drone uses a camera and LiDAR-based vision system to estimate its pose relative to the power line, though the algorithm's performance depends on adequate lighting. Building on this work, the paper \parencite{Hamelin_2019} presented a vision-based feedback control method for the LineDrone hybrid drone, employing a discrete-time controller. Real outdoor tests demonstrated excellent performance; however, the perception system still faced limitations due to the camera's field of view and dependency on daylight. Several other researches have been done concerning the landing of a drone on a power line using various methods, including top and bottom grasping perching \parencite{Popek_2018} \parencite{Iversen_2020} \parencite{Meng_2022}. This method of bottom perching works well to ensure the stability of a small drone on a cable with grippers and can perform drone self-recharging \parencite{Iversen2021}, but is limited because it does not allow movement on the power line. 

Landing envelopes have been extensively studied and applied in a variety of contexts beyond power line inspection, such as rooftop landings, wall landings, landing on moving objects, and even in aerospace applications like helicopter landings or spacecraft touchdowns. In the case of rooftop landings \parencite{Bass_2022} and wall landing \parencite{Mehanovic_2017}, strategies have been devised for safe and reliable landings on flat and inclined surfaces, considering factors like wind disturbances, vertical impact velocity, and accurate localization. For moving objects such as vehicles or ships, or even SpaceX's Falcon 9 rocket \parencite{blackmore2016autonomous} landing on a drone ship, the challenge is greater due to the need for precise relative localization and adaptive control strategies that can accommodate the dynamic nature of the landing platform \parencite{Keipour_2021}. Even the Mars Rovers' landing sequence \parencite{Holzmann_2013} is a complex process where a landing envelope must be carefully calculated and executed. These diverse landing scenarios underscore the broad applicability of landing envelope concepts and the significance of developing context-specific landing strategies.

In the field of robotics and aerospace, the Monte Carlo method is a powerful computational technique used for probabilistic reasoning, estimation, and decision-making. This approach involves simulating numerous scenarios by randomly sampling possible outcomes, allowing robots to better understand and predict their environment online or offline. Monte Carlo algorithms, such as particle filters \parencite{Gordon_1993} and Monte Carlo Localization (MCL) \parencite{Dellaert_1999} \parencite{Akai_2022}, are widely applied in robotics for tasks like localization \parencite{Montemerlo_2003}, mapping \parencite{Alberto_2022}, and motion planning \parencite{Hu_2021}. By incorporating uncertainty and considering a range of potential outcomes, the Monte Carlo method enables robots to make more informed decisions and adapt to complex, real-world situations with greater reliability and accuracy.

In summary, this state-of-the-art section has provided an overview of drone landing for power line inspection, discussing drones, sensors, landing research, landing envelopes, and the application of Monte Carlo methods in robotics. This review lays the groundwork for the development of an improved landing strategy, incorporating the latest advancements in drone technology, sensor systems, and computational methods.
    \section{Methodology}\label{sec:Methodology}

\begin{figure}[H]
    \centering
    \includegraphics[width=0.8\textwidth]{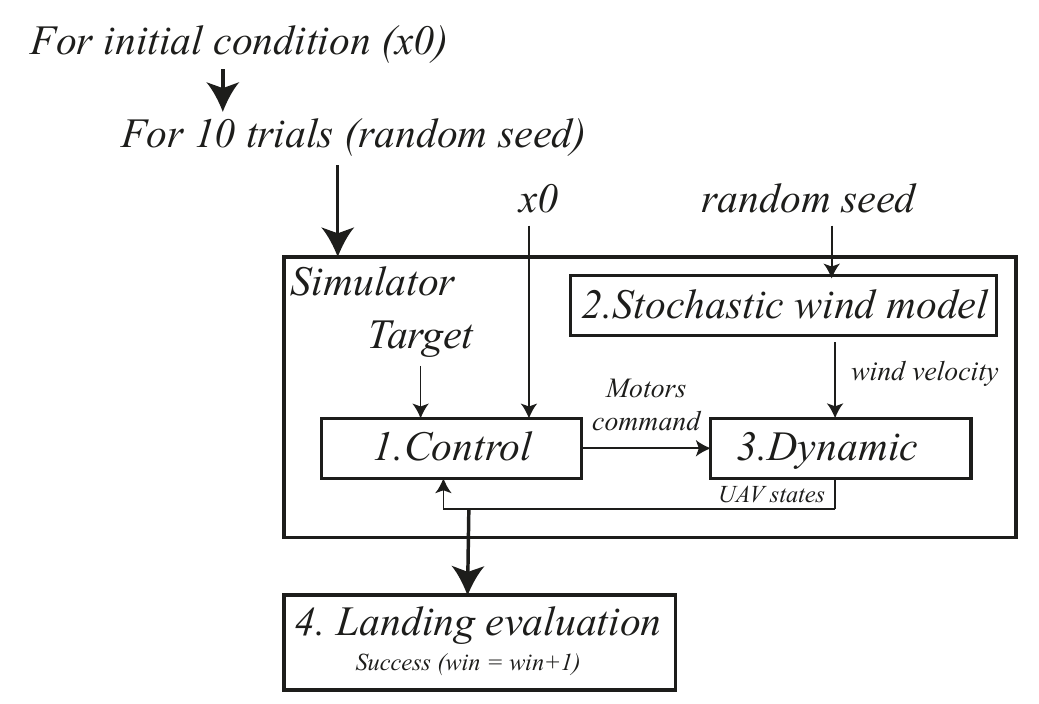}
    \caption{Methodology overview}
    \label{fig:overview}
\end{figure}

To determine the landing envelope of the NADILE under various wind conditions, we utilize a tailored approach. This approach involves employing a simulation model of the drone and conducting a series of direct landings to accurately calculate the landing envelope. Figure \ref{fig:overview} demonstrates the approach employed to establish the landing envelope under windy conditions. A Monte Carlo method was developed to analyze the probability of successful landing based on initial drone states, offering valuable insights for refining the landing strategy. We perform simulations for each initial state vector \(x_0\) , encompassing the drone's position, orientation, and velocity at \(t=0\). For each \(x_0\), ten trials are conducted, each with a randomly generated wind speed profile centered around 5, 10, 15, and 20 km/h in the most unfavorable lateral conditions. Subsequently, a direct landing of the drone is simulated and evaluated using a gain function.

The next sections describe the components of the control strategies in subsection \ref{sec:control}, the stochastic wind model in subsection \ref{sec:wind model}, the simulation model in subsection \ref{sec:dynamic model} and the methodology for landing evaluation in subsection \ref{sec:landing eval}.

\subsection{Landing strategies}

\begin{figure}[ht]
    \centering
    \includegraphics[width=0.8\textwidth]{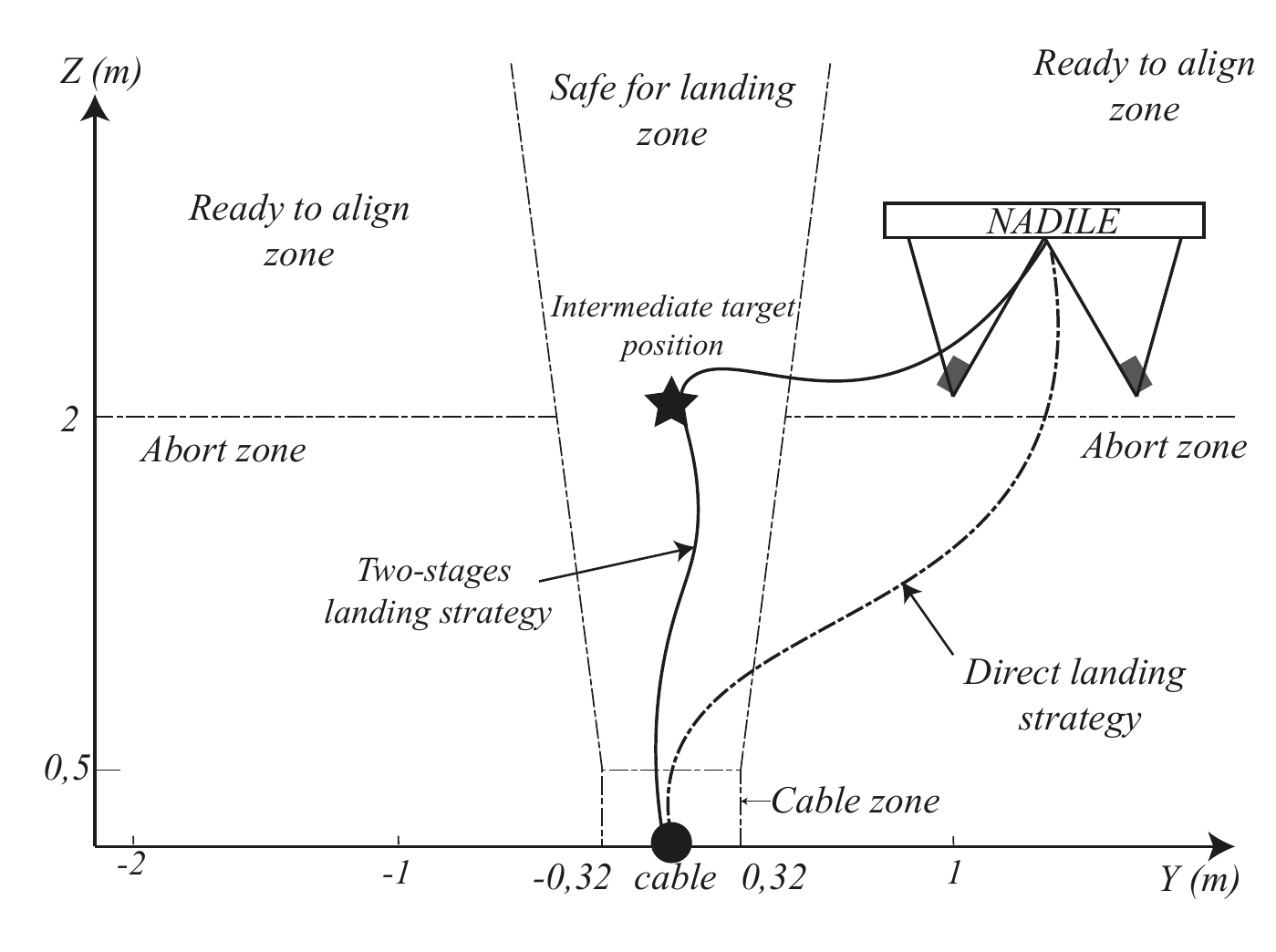}
    \caption{Landing zones around cable with both landing strategies (TSLS and DLS)}
    \label{fig:LandStrats}
\end{figure}

Two landing strategies has been evaluated and implemented in this study, the direct landing and the two-stage landing. They are illustrated in Figure \ref{fig:LandStrats}.

In the direct landing strategy (DLS), the drone aims to land directly on the cable, targeting the cable itself. While this method might be quicker, ensuring a safe landing could be more difficult, as the drone might not have sufficient time to adjust its position, orientation, and speed before making contact with the cable. Moreover, the drone's structure, depicted in Figure \ref{fig:LineDroneNadile}, consists of a substantial aluminum enclosure standing at a height of 1 meter. This particular design characteristic presents a potential obstacle when aiming to establish a secure and reliable approach from various starting positions.

\begin{figure}[ht]
    \centering
    \includegraphics[width=0.8\textwidth]{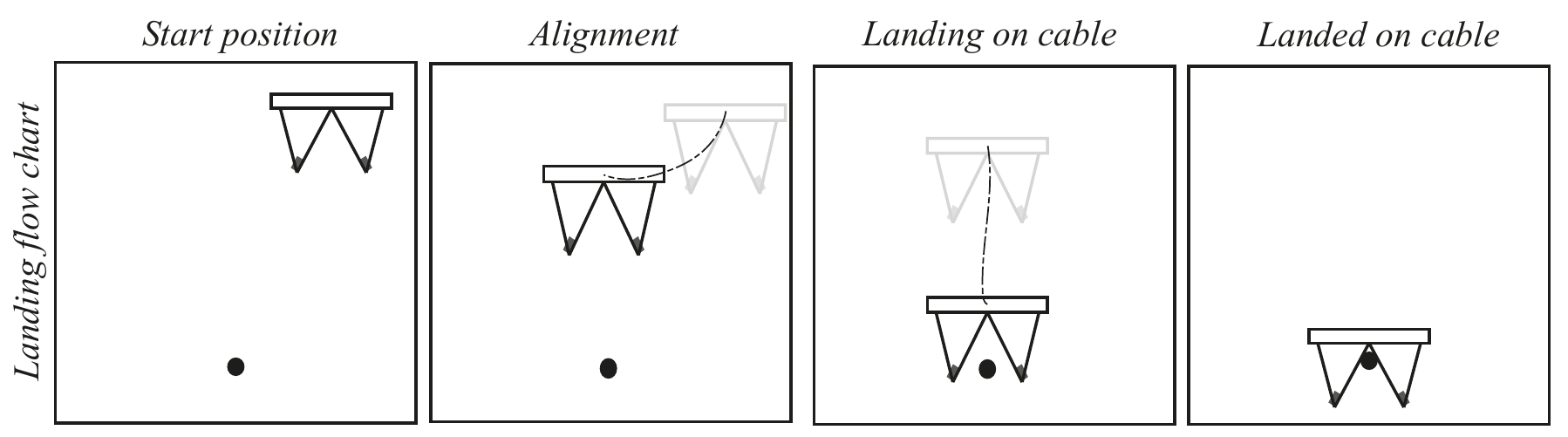}
    \caption{Landing flow chart from start position to land on cable}
    \label{fig:flow}
\end{figure}

The two-stage landing strategy (TSLS) is a safer approach that consists of two distinct phases: alignment and landing. In the alignment phase, the drone first maneuvers itself to an intermediate target position above the cable, ensuring that its position, orientation, and speed are within acceptable bounds. As will be discussed later, the alignment phase significantly broadens the region where the pilot can commence the landing phase (including areas ready for line-up and safe landing zones), in comparison to the Direct Landing Strategy (DLS), which might traverse undesirable zones (e.g., abort zone). Landing zones illustrated in Figure \ref{fig:LandStrats} are regions that have been qualitatively defined through experience and practical knowledge, providing a suitable area for drone to land safely during their operations. The results of this study aim to confirm and refine the geometry of these landing zones, providing a more systematic and data-driven approach to identifying optimal landing areas. As illustrated in Figure \ref{fig:flow}, once the drone is properly aligned, it proceeds to the landing phase, where it descends and makes contact with the cable. This strategy provides more opportunities for error correction and adjustments during the alignment phase, resulting in a safer and more controlled landing. 

The TSLS is governed by a landing state machine, which continually tracks the drone's states during the landing process. This state machine is tasked with ensuring the drone stays within acceptable boundaries regarding position, orientation, and velocity as it moves between the alignment and landing phases. As the drone progresses through the landing sequence, the state machine confirms that all required conditions are fulfilled before permitting the drone to proceed to the next state. 


\subsection{Alignment and landing controllers}

\begin{figure}[H]
    \centering
    \includegraphics[width=1\textwidth]{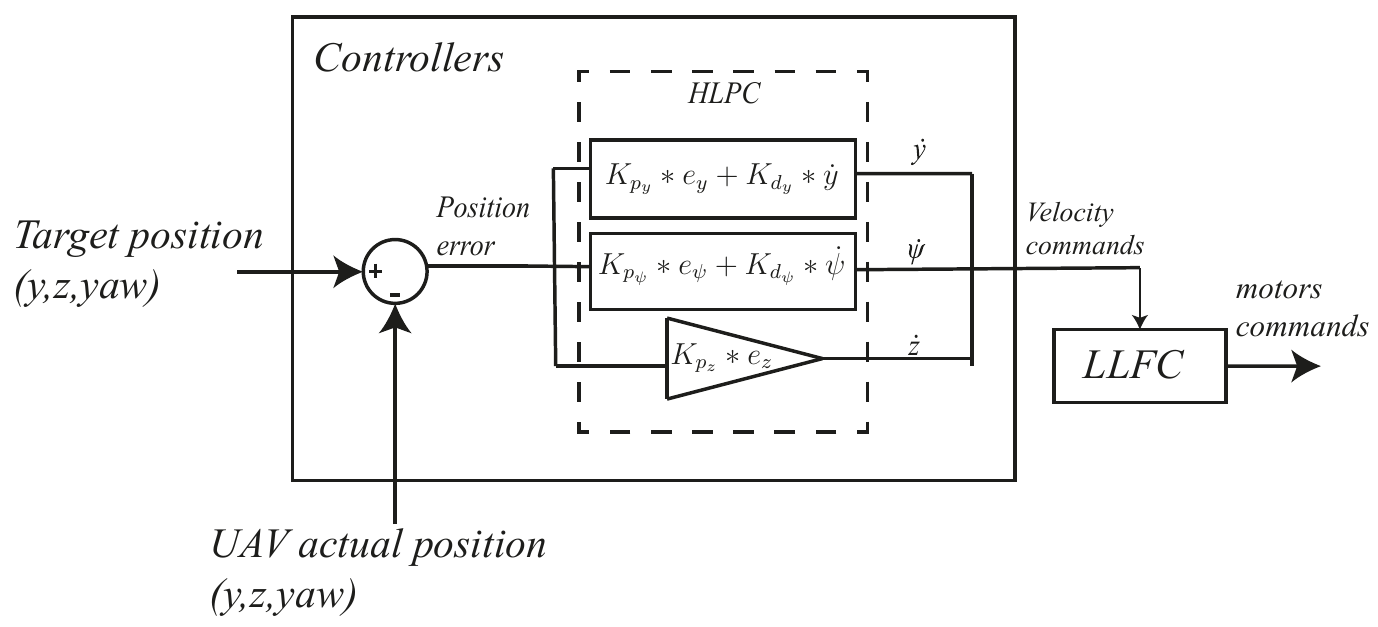}
    \caption{The custom HLPC takes position errors as input and provides velocity commands to the LLFC.}
    \label{fig:landcontroller}
\end{figure}

In the custom configuration of the NADILE, the drone's position relative to the cable during landing sequence is managed by a high-level position controller (HLPC), while the drone's stability is maintained by the low-level flight controller (LLFC) using the autopilot internal control loops. The HLPC comprises three decoupled position controllers (y, z, and yaw), each responsible for moving the drone toward the target position along their respective axes.The pilot and the LLFC handle the x-axis which is parallel to the conductor. Lateral (y) and yaw errors are managed by two proportional-derivative (PD) controllers, while a proportional (P) controller handles altitude control (z). From the HLPC perspective, the system being controlled is at least a type 1 system, i.e. it has at least one pole at the origin in the continuous-time domain. Hence, the steady-state error will naturally tend toward zero even if the HLPC has no integrator.  The HLPC configuration is illustrated in Figure \ref{fig:landcontroller}. In this configuration, \(K_{p_y}, K_{p_{\psi}}\) and \(K_{p_z}\) are proportional gains that multiply the errors between the target and actual drone's pose, while \(K_{d_y}\) and \(K_{d_{\psi}}\) are the derivative gains that multiply the time derivatives of the errors. The target is obtained using a LiDAR-based perception system described in \parencite{Périard_2022}, which is out of the scope of this paper. The landing phase also employs the same HLPC to ensure proper drone alignment. The HLPC generates velocity commands that are sent to the LLFC in ArduPilot \parencite{copter} via Mavlink protocol \parencite{mavlink}.

In summary, the alignment and landing controllers, landing strategies, and landing state machine are presented, and their performance will be evaluated for multiple control gains and wind speed. These various mechanisms were incorporated into simulations to assess landing envelopes under a multitude of parameters. This understanding will facilitate the development of more efficient, safer, and more reliable landing procedures for a variety of inspection, maintenance, and other applications.
\label{sec:control}
 \subsection{Stochastic wind model}

A vital element of the Monte Carlo model is accurately representing wind variability during the landing process. To simulate appropriate wind profiles, the Von Kármán wind spectrum \parencite{Abrous2015} was used for each simulation trial.
This Von Kármán wind spectrum describes the wind speed at a specific location and time using the following general equation:

\begin{equation}
V(f) = V_0 * f^{(\frac{-5}{3})}
\end{equation}

\begin{figure}[ht]
    \centering
    \includegraphics[width = 0.5\textwidth]{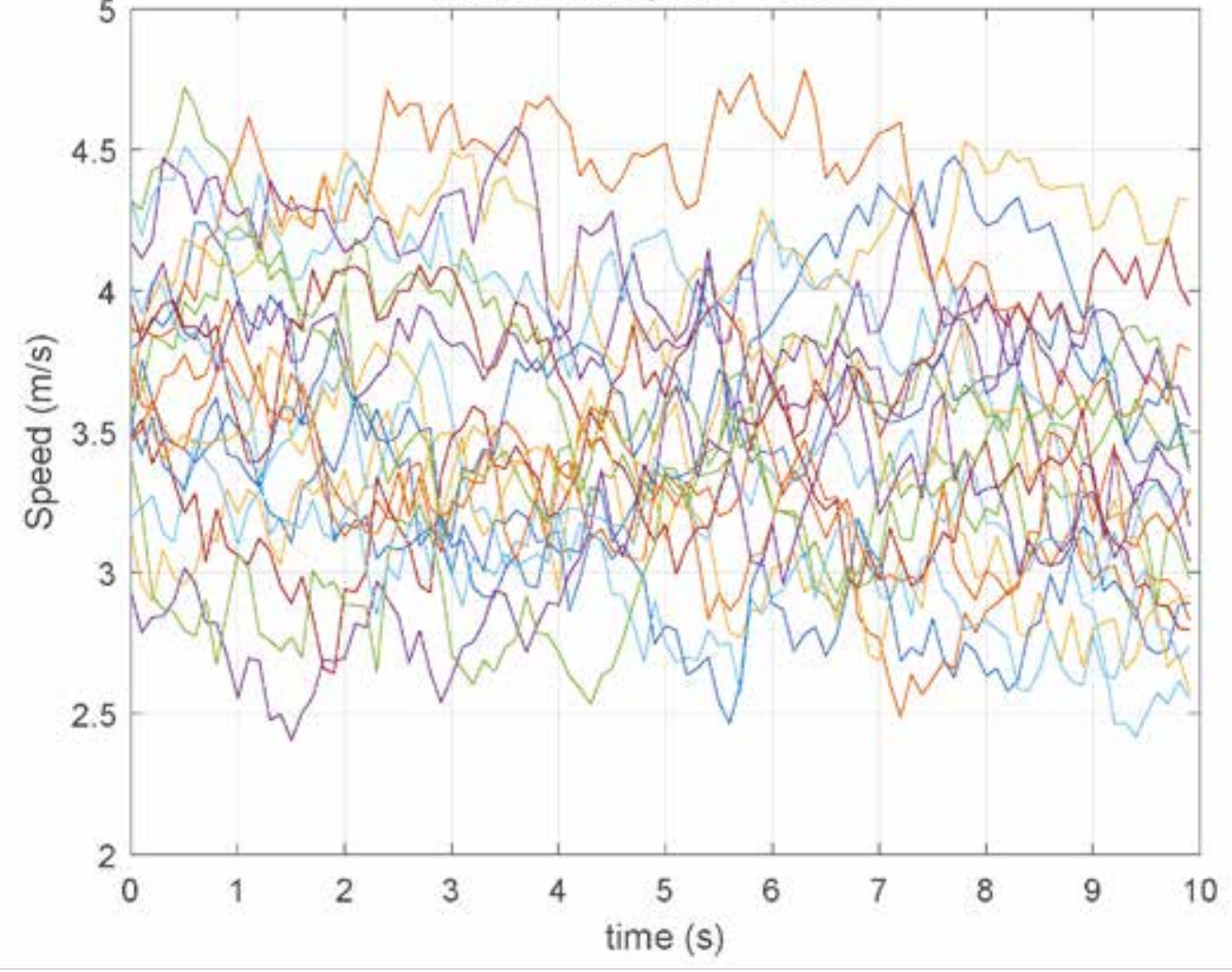}
    \caption{3.5 m/s wind samples generated by the Von Kármán wind generator}
    \label{fig:wind_sample}
\end{figure}

In this equation, \(V(f)\) represents the wind speed for a given frequency \(f\), \(V_0\) is a scaling factor denoting the average wind speed at the given location, and the exponent \(-5/3\) is known as the von Kármán exponent. This exponent describes the relationship between the frequency of the wind fluctuations and their intensity, with higher frequencies corresponding to smaller-scale turbulence. This spectrum offers a mathematical framework for generating the distribution of wind velocities. The random seed of the Monte Carlo method is a random wind profile defined by this wind generator. The average wind speeds are fixed, but the random profile allows for the evaluation of landings under varying wind speeds over time as illustrated in Figure \ref{fig:wind_sample}. This enabled the Monte Carlo model to account for the inherent variability and turbulence encountered by the NADILE during the landing process. By simulating a range of wind conditions, the study aimed to provide a comprehensive analysis of the landing envelope and identify strategies for safe and reliable landings on high-voltage electric cables.

 \label{sec:wind model}
\subsection{Dynamic}\label{sec:model drone}
\begin{figure}[ht]
\centering
\includegraphics[width=1\textwidth]{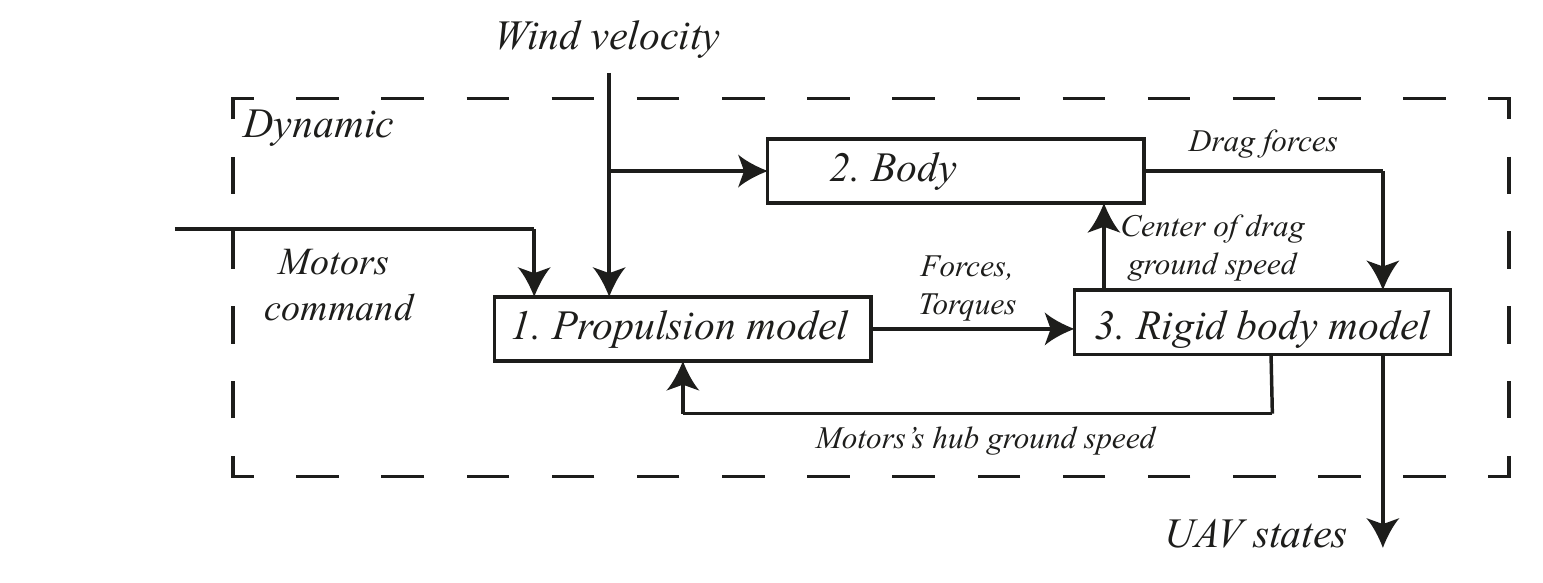}
\caption{Model flow chart for dynamic simulations}
\label{fig:model_flow_chart}
\end{figure}

To accurately evaluate the NADILE's landing envelope in various wind conditions, a realistic drone dynamic was necessary. Indeed, the wind generate  additional aerodynamics forces on the drone fuselage, but mainly from the interactions between the wind and propellers. If these additional forces are ignored, it can lead to an overestimation of the drone's landing envelope in high wind. Thus, in addition to the usual motors' command, the drone model developed takes as input the wind velocity to compute the additional aerodynamics forces  to calculate the main states of the drone (i.e., 6-DOF position and velocity). The links between the main processes of the drone's dynamic model are presented in Figure \ref{fig:model_flow_chart}.

This section presents a realistic propulsion model that could take into account the coaxial propellers interaction and the airspeed effects using wind tunnel tests to create an empirical propulsion model. Using the airspeed (wind speed minus the drone's ground speed) and the commands from the flight controller, the propulsion model output the total forces and moments generated by the propellers. For the aerodynamics forces generated by the fuselage, due its geometric complexity, CFD analysis were done to obtain the the drone's drag forces function of the airspeed. The calculated forces and moments are then use in a 3D rigid body model to calculate the drone's dynamic and to obtain the states that will be used by the flight controller.

\subsubsection{Propulsion model}
\label{sec:propulsion_model}

\begin{figure}[ht]
\centering
\includegraphics[width=1\textwidth]{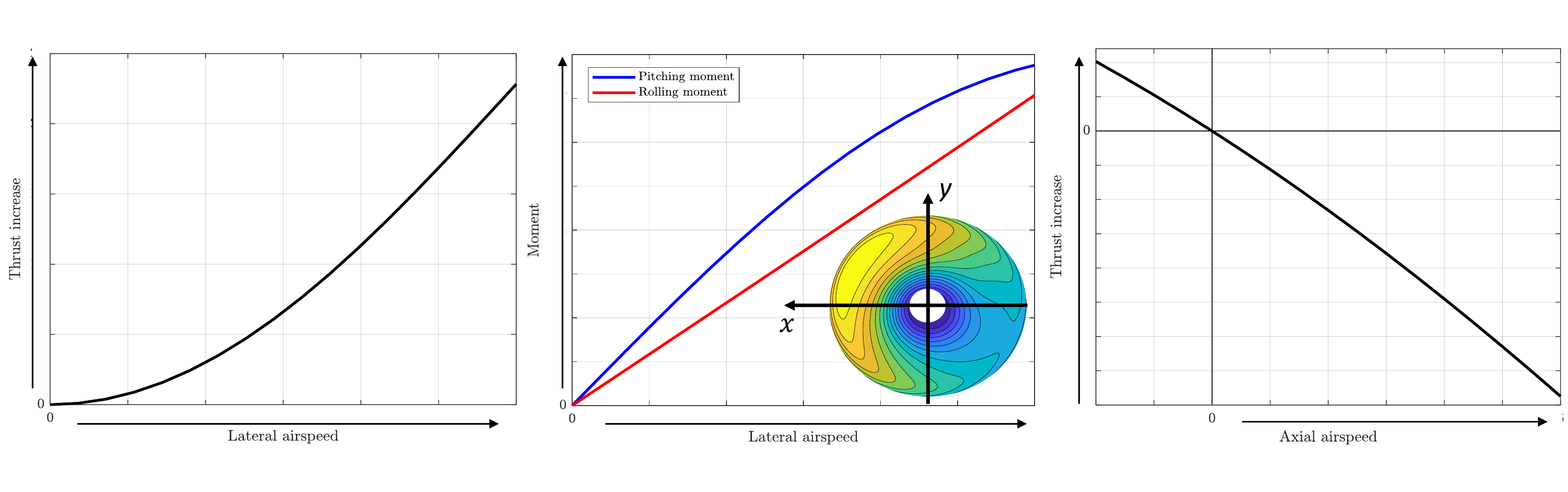}
\caption{Aerodynamic effects of airspeed on a multirotor drone propeller considering a fixed rpm. From left to right: the increase in thrust with changes in lateral airspeed, the increase in the pitching and rolling moments with changes in lateral airspeed and the change (positive or negative) of the thrust depending of the sign of the axial airspeed.}
\label{fig:aero_effects}
\end{figure}

Propulsion forces/torques are often modelled as simple function of the motors' rotation speed \parencite{simple_propulsion}. However, to capture the effect of the wind on the NADILE, more advanced models that consider the effects of airspeed are required. Lateral and axial wind affect the propeller's forces in three main ways \parencite{leishman_principles_2016}, as presented in Figure.\ref{fig:aero_effects}.

There many effect caused by lateral and axial airspeed on the propellers. The first effect is the increase in the thrust caused by the increase of the lateral airspeed.This cause a reduction of the induced speed in the rotor resulting in a higher angle of attack for the same rotational speed. The second effect caused by lateral airspeed is the creation of rolling and pitching moments on the rotor hub due to the asymmetric lift distribution on the rotor. For example, take the rotor illustrated in Figure \ref{fig:aero_effects}. The rotor hub airspeed is in the +X axis direction, and the propeller rotates counterclockwise. The advancing blade experiences a higher airspeed than the retreating one, creating a positive pitching moment (Y axis) and negative rolling moment (X axis) about the propeller's hub. Since the propellers are rigid for a multirotor drone (no hinges), the rolling and pitching moments of each rotor are transmitted to the drone's body \parencite{leishman_principles_2016}. The third effect is the increase or decrease of thrust due to a change in the axial airspeed, which happens during climb and descent, but also during rolling and pitching maneuvers. They effectively act as a dampener where a positive vertical speed reduces the thrust and a negative speed increases the thrust for the same rotational speed.  

\begin{figure}[ht]
\centering
\includegraphics[width=1\textwidth]{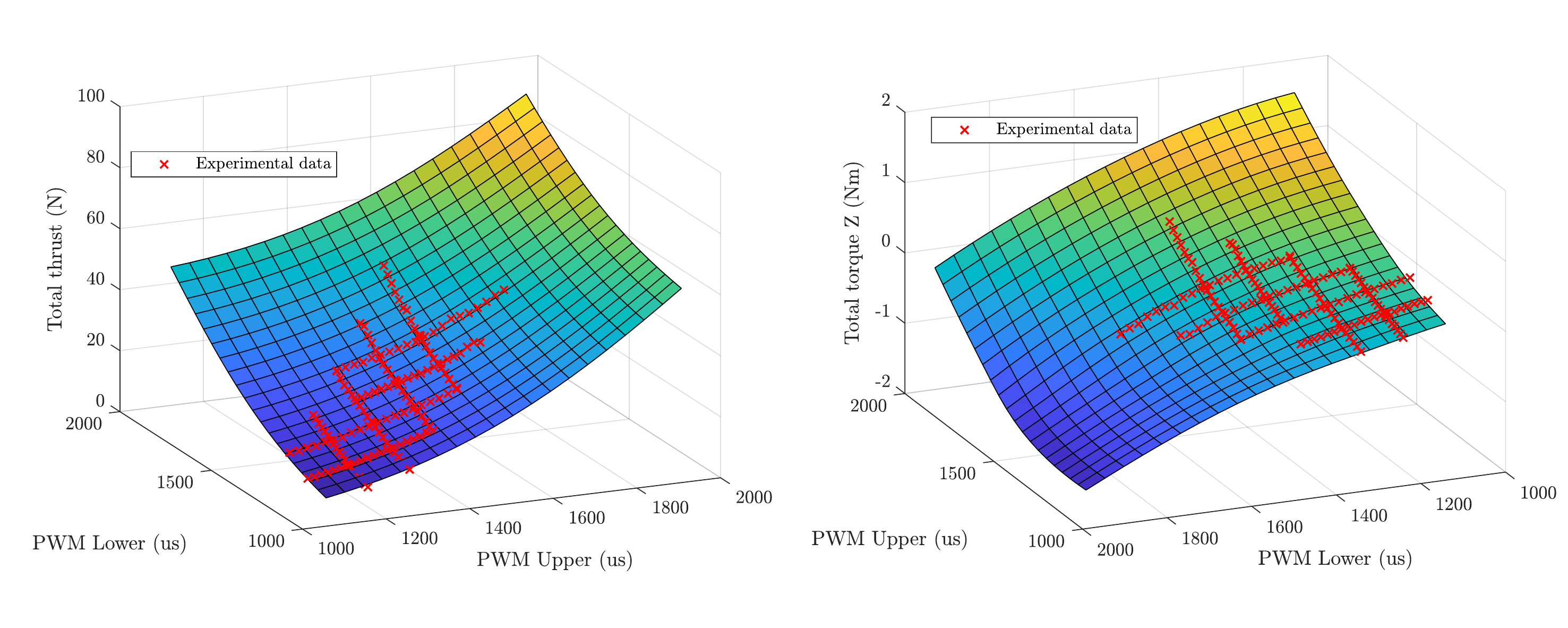}
\caption{Total thrust (left) and total Z axis (shaft) torque  vs motors' command generated by using the trained functions compared to the experimental data in wind tunnel. The graphs shown were obtained for a 0~m/s airspeed and 0\textdegree~angle of attack.}
\label{fig:exp_prop_fit}
\end{figure}

The drone uses four pairs of coaxial propellers, which considerably increase the complexity of the aerodynamics effects between the upper and lower propellers\parencite{coax_rotor_analytic}. To create a high-fidelity model while avoiding CFD calculations\parencite{coax_rotor_cfd}, the drone's coaxial propulsion was tested on two \textit{Tyto Robotics} Series 1780 Thrust Stand \parencite{tyto} in a coaxial configuration. The setup was placed in a wind tunnel where multiple runs were done while changing one of the four main parameters (upper motor RPM, lower motor RPM, wind speed, and pitching angle) one at a time. The motor's command range tested was from 1200~us to 1700~us for both motors, 0~m/s to 10~m/s for the wind speed and -10\textdegree~to 10\textdegree~for the angle of attack. Using the experimental data, an empirical propulsion model was created that takes as input the upper and lower motor RPM and the relative airspeed magnitude and direction to calculate the resulting forces and torques of each coaxial pair. The model was created using MATLAB Neural Network Training function using the Bayesian Regularization training. The model was then validated against experimental data that was not used in the training process. The training is finally used to generate functions that calculate the forces/torques in the main axis that will be used in the dynamic model. An example of the force/torques obtain using the trained function is presented in Figure \ref{fig:exp_prop_fit} for for a 0~m/s airspeed and 0\textdegree~angle of attack. We can observe the experimental data superposed correctly over the trained model (surface) and allows extrapolation the experimental data to cover to entire motors' range.

\subsubsection{Structural Drag}
\label{sec:body_drag}
\begin{figure}[ht]
\centering
\includegraphics[width=1\textwidth]{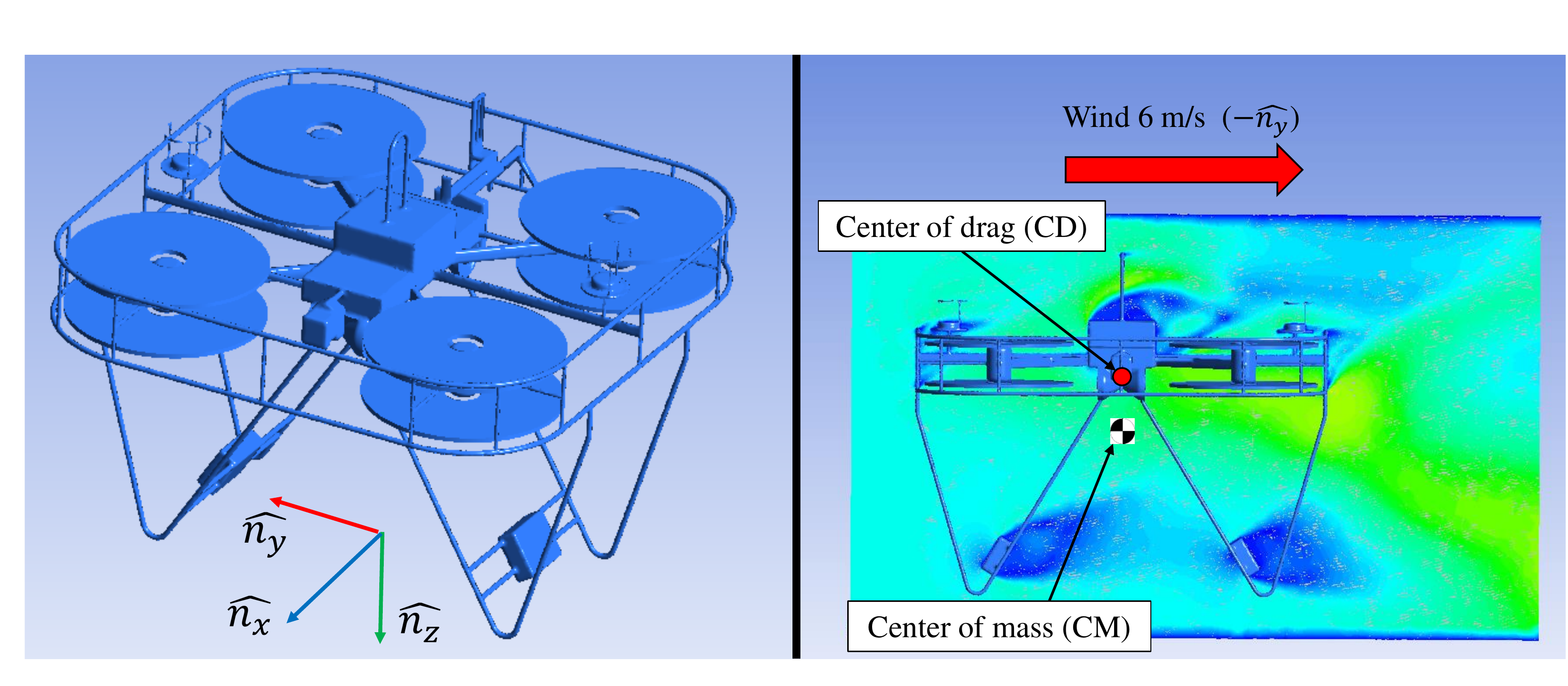}
\caption{Computer fluid analysis of the drone structure drag. On the left, is the simplified 3D model used for the analysis and on the right, is a section view with the wind speed contour around the drone for a 6~m/s wind.}
\label{fig:cfd_linedrone}
\end{figure}

 The design constraints of the  NADILE's mission resulted in a unique structure with a large aluminum cage to protect delicate components against electrical discharges and physical contact. Also, the need to be passively stable when landing on the power lines required the center of mass to be low, which was achieved by placing the batteries at the base of the legs. The large cage and batteries at the bottom are an important source of drag, and CFD analysis was performed to obtain an estimate of the NADILE's complex structure drag, as presented in Figure \ref{fig:cfd_linedrone}. We can observe that the batteries at the base of the cage and the central structure cause significant disturbances in the airflow, resulting in a high drag force. 

Different drone inclinations at various airspeeds were tested to determine the \textit{Equivalent Flat Plate} area (Drag coeffcient of 1), which will be used to calculate the drag at different airspeed in the three primary directions. The analysis yielded an equivalent falt plat area of 0.512~$\mathrm{m^2}$. Another important outcome of the CFD analysis is the position of the center of drag (CD). The Center of Drag (CD), where all aerodynamic forces on the drone's frame converge, determines the stability of the drone under high airspeed \parencite{effect_CD_position}. In our case, due to the unique configuration, the CD is above the Center of Mass (as seen in Figure \ref{fig:cfd_linedrone} right), making lateral airspeed destabilizing for the drone.

\subsubsection{Rigid body model}

\begin{figure}[ht]
\centering
\includegraphics[width=0.8\textwidth]{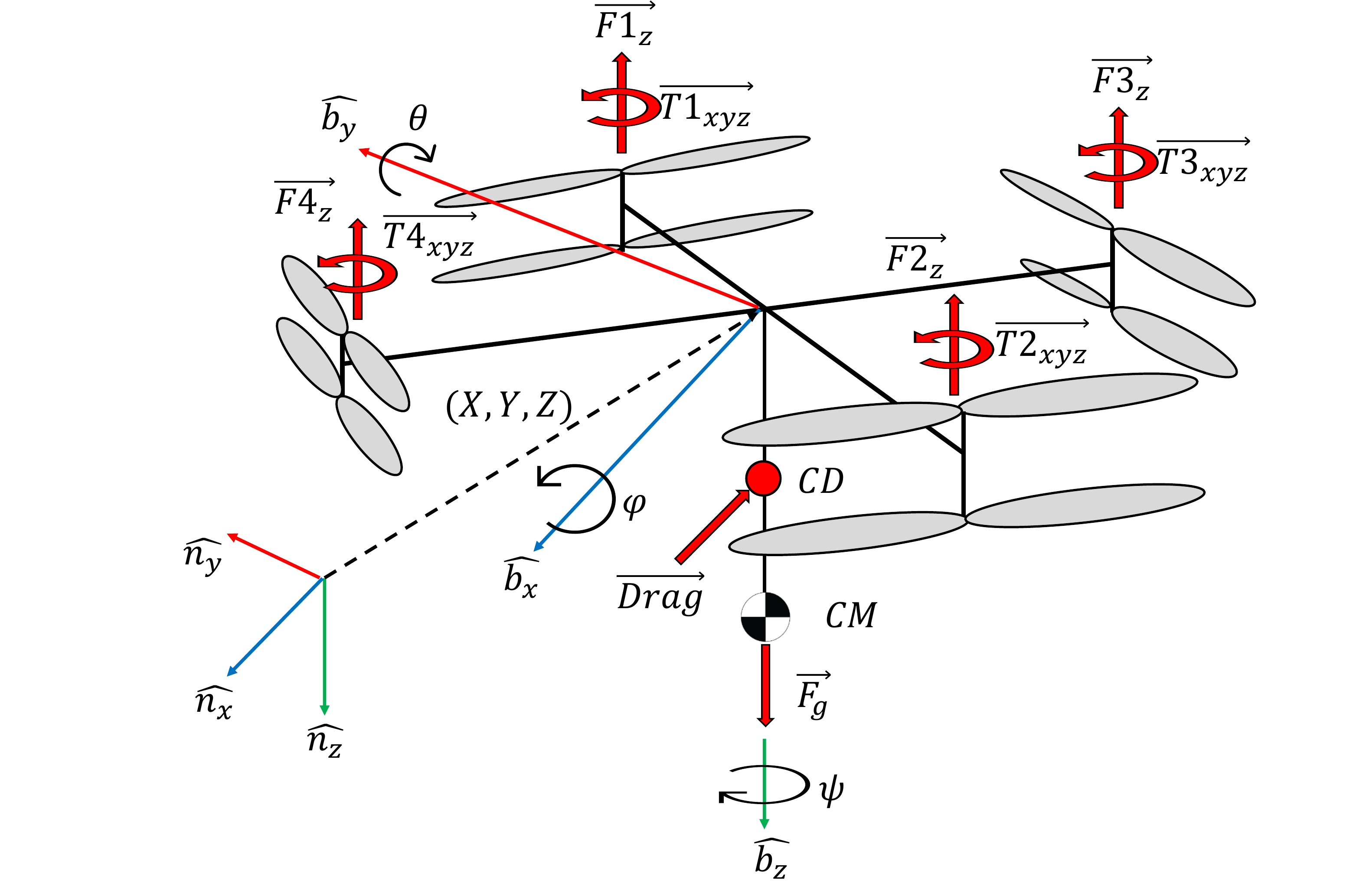}
\caption{Free body model of the drone and applied forces/torques for the rigid body dynamic. }
\label{fig:dcl_nadile}
\end{figure}

The forces and torques defined in the previous sections are used to calculate the drone's dynamics. The drone's structure is approximated as a single rigid body. The free diagram model used to define the dynamics is presented in Figure \ref{fig:dcl_nadile}.

The model uses the two standard frames generally used to model a drone's dynamic; the inertial frame \textit{N} is fixed on the ground, and the body frame \textit{B} is attached to the drone. These frames use the conventional north-east-down (NED) for the main directions. The mass and inertia of the rigid body were measured experimentally with the NADILE prototype. The drone's mass is 20~kg with X,Y and Z axis inertia in the body frame being respectively 3.52, 3.31 and 3.84~$\mathrm{kg.m^2}$. For the applied forces and moments, each coaxial pair is considered as one entity, as defined in Section \ref{sec:propulsion_model}, and applies a force in the body frame negative Z-axis with torques in the body X-Y-Z axes. The body drag forces presented in Section \ref{sec:body_drag} are applied at the center of drag (CD).The drone's physical properties and the forces/moments are then used to solve the differential equation of motion in translation and rotation for the 3 main axis. The solved equations are then used at each time step of the simulation to update the new states bases on the previous ones and the forces/moments applied.




\label{sec:dynamic model}
\subsection{Landing envelope evaluation}

In this section, a statistical probability approach based on the Monte Carlo algorithm is used to determine the landing envelope of the NADILE. The Monte Carlo simulation method, as shown in Figure \ref{fig:overview}, is a statistical technique that involves generating numerous random samples (also known as "trials") of a system and using them to approximate the probability of specific outcomes. The simulation used the MatLab and Simulink implementation of the drone simulation model based on Section \ref{sec:model drone}. Moreover, to simulate realistic wind conditions, a wind generator utilizing the Von Karman algorithm was employed. This method, as detailed in Section \ref{sec:wind model}, generated wind samples with varying average wind speeds within the manual operational range (ranging from 0 km/h to 20 km/h). The wind samples, illustrated in Figure \ref{fig:wind_sample}, were generated with a standard deviation of 3.6 km/h to accurately simulate wind speed fluctuations.

Ensuring the safety and reliability of the drone during its high-voltage cable landing operations heavily relies on meeting specific landing conditions. Successful landing is determined by the success function, which validates compliance with the following defined conditions, with the tolerances indicated in the table \ref{tab:tol_table}.

\begin{figure}[H]
    \centering
    \includegraphics[width = 0.6\textwidth]{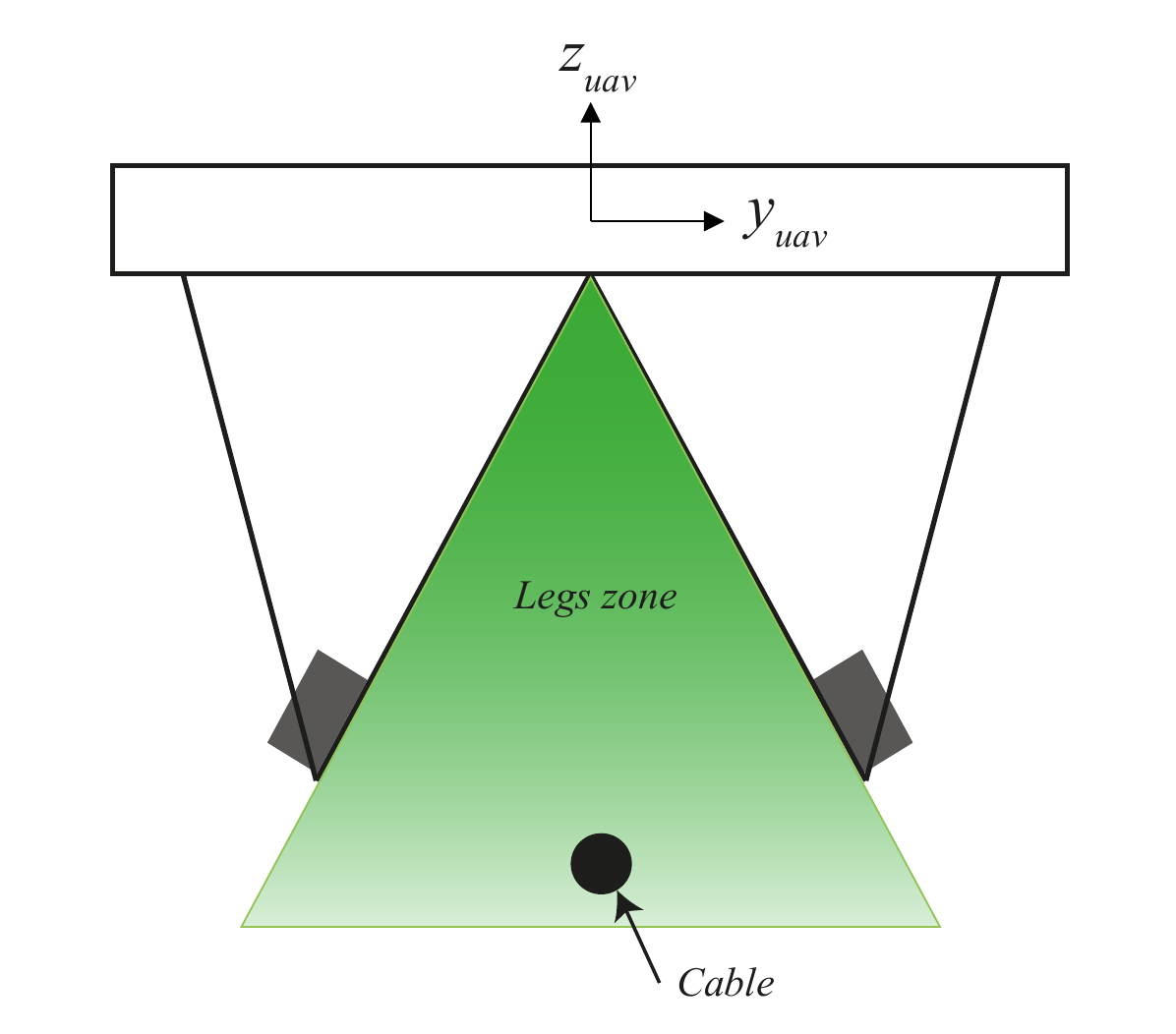}
    \caption{Legs zone for cable validation}
    \label{fig:Legszone}
\end{figure}

\begin{enumerate}
    \item The cable is located in the drone's legs zone as Figure \ref{fig:Legszone}
    \item The absolute roll angle of the drone \(|\phi|\) is below  \(|\phi_{tol}|\) .
    \item The absolute roll velocity  \(|\dot{\phi}|\) is below  \(\dot{\phi}_{tol}\).
    \item The velocity vector norm \(|\vec{v}|\) is below \(|\vec{v}|_{tol}\).
    \item The direction of the velocity vector  \(\vec{v}\) is directed toward the cable within some tolerance \(\vec{v}_{tol}\) 
    \item The absolute yaw angle between the drone and cable \(|\Delta\psi|\) is below \(|\Delta\psi|_{tol}\)
    \item The absolute yaw velocity \(|\dot{\psi}|\) is below \(|\dot{\psi}|_{tol}\).
\end{enumerate}

\begin{table}[h]
    \centering
    \begin{tabular}{c|c|c}
        Variable & Maximum & Unit \\
        \hline
       \(|\phi|_{tol}\) & 0.08 & rad\\
       \(|\dot{\phi}|_{tol}\)  & 0.08 & rad/s \\
       \(|\vec{v}|_{tol}\) & 0.2 & m/s\\
       \(\vec{v}_{tol}\) & \([\pm0.5,-0.2]\) & m \\
       \(|\Delta\psi|_{tol}\) & 0.1 & rad\\
       \(|\dot{\psi}|_{tol}\) & 0.1 & rad/s\\
    \end{tabular}
    \caption{Tolerance table}
    \label{tab:tol_table}
\end{table}

To consider a landing successful, the drone's states must meet specific criteria when entering the cable zone. These criteria are derived from the need to control safe landing conditions. It is necessary to control the speed and direction of the drone towards the cable prior to contact. The maximum allowable speed is determined by the theoretical descent speed command of 0.2 m/s. It is also important to verify the tilt angle within the range of (\(\pm 5deg\) degrees and the angular velocity within (\(0.5 deg/s\) to ensure pitch stability during landing. Finally, it is crucial to ensure that the cable is between the drone's legs to guarantee a successful landing.
If the win function is asserted, i.e. all conditions above are satisfied, a successful landing can be attributed to the initial position \(x_0\). The Monte Carlo evaluation was conducted using starting points within a lateral range of (-1.5, 1.5m) and an altitude range of (1.5 to 2.5m).

\label{sec:landing eval}

    \section{Results} \label{sec: Results}
In this section, we present the results of our study, which are organized into three main areas: gain evaluation, landing envelope evaluation, and landing strategies comparison. First, we discuss the gain evaluation process, which determines the optimal control gains for the high-level position controller (HLPC) during the alignment phase. Next, we assess the landing envelope for the NADILE under various wind conditions, providing insights into the drone's performance and limitations during the landing process. Finally, we compare the direct landing strategy (DLS) and the two-stage landing strategy (TSLS) to highlight the advantages and drawbacks of each approach in terms of success rate and overall safety.


\subsection{Gains evaluation results}

\begin{figure}[h]
    \centering
    \includegraphics[width=0.9\textwidth]{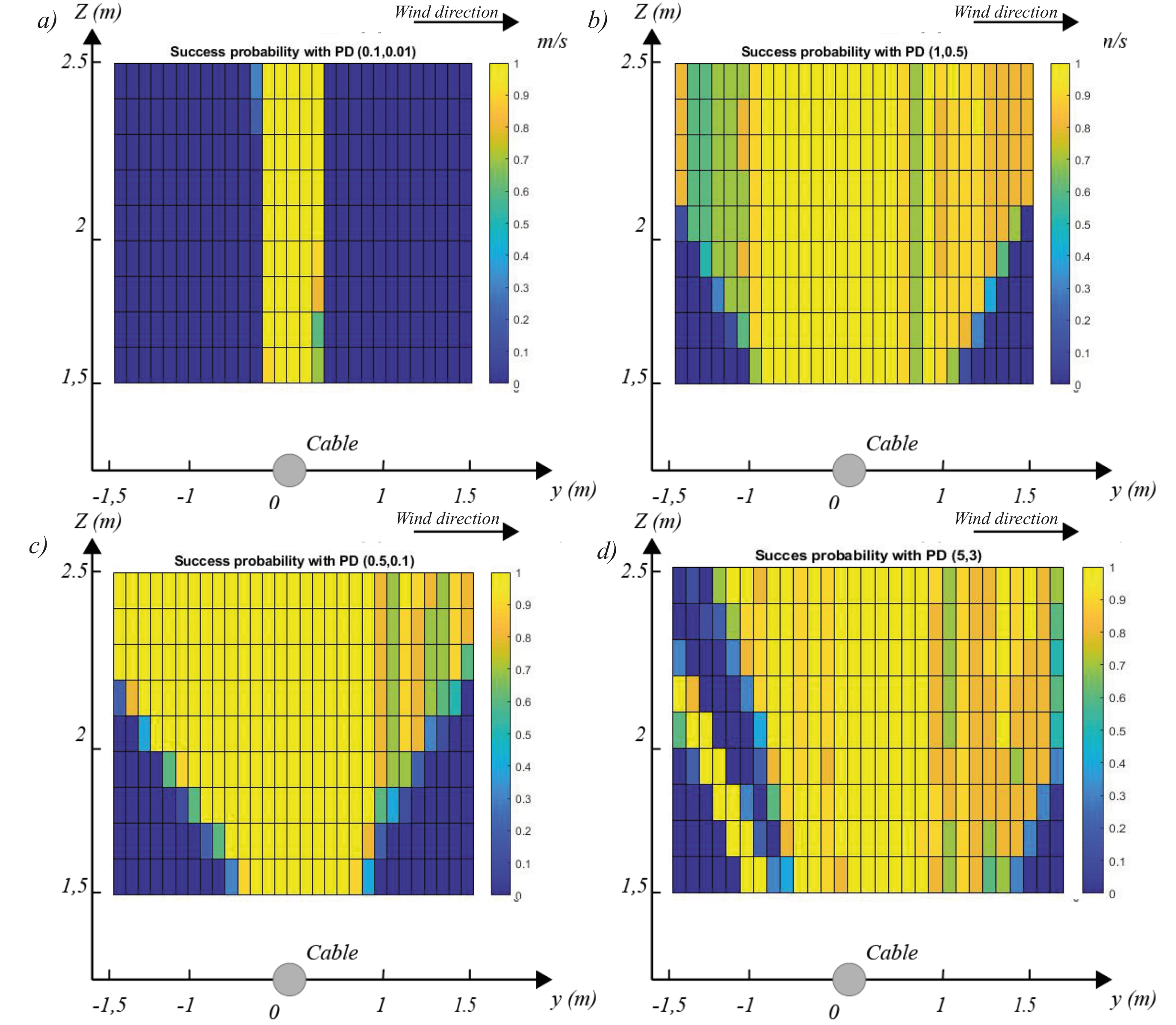}
    \caption{The success maps from the evaluation of controller gains are as follows: in a), low gains result in a very restricted success zone. In b), slightly increasing the PD gains to (1,0.5) slightly widens the success zone. In c), the success zone is considerably broad with PD gains of (0.5,0.1). However, in d), high PD gains of (5,3) reduce the chances of success.}
    \label{fig:MapsGains}
\end{figure}
\begin{figure}[h]
    \centering
    \includegraphics[width=0.8\textwidth]{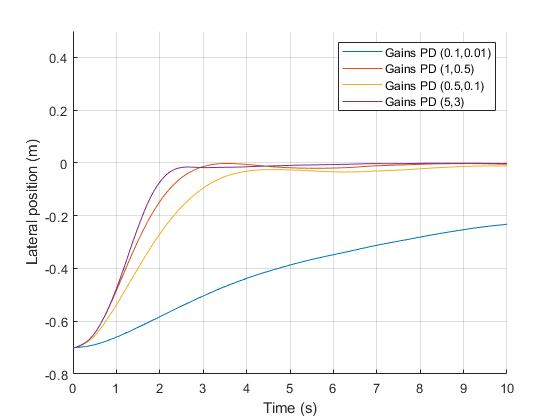}
    \caption{Time response of the lateral position of the drone with respect to the cable for PD couples gains.}
    \label{fig:stepresponse}
\end{figure}

In this section, we undertake an appraisal of the PD gains for the high-level position controller. We tested the \(K_p\) and \(K_d\) gains of the alignment PD controller within the ranges of \([0.1:0.5:5]\) and\([0.01:0.1:3]\) respectively, aiming to identify the optimal couple of PD gains for the alignment controller. Securing gains that deliver the largest possible success zone while maintaining an alignment time under 5 seconds assures a successful landing. The altitude and yaw controller gains were established empirically through simulations and confirmed through experimental trials, since the drone's geometry permits a larger margin of error in altitude and yaw.
The gains evaluation results are shown in Figure \ref{fig:MapsGains}. The figure \ref{fig:stepresponse} depicts the temporal responses of different PD gain pairs under a wind speed of 10km/h. The PD gain pair (0.5,0.1) demonstrates a smooth and consistent alignment with stabilization achieved in under 5 seconds. Maintaining a low \(K_p\) reduces the drone's abrupt accelerations during movements, thereby enhancing the likelihood of successful landing.

As illustrated in Figure \ref{fig:MapsGains}, low controller gains \(a)\) resulted in a landing envelope limited to the zone directly above the cable, while high controller gains \(d)\) demonstrate instability in the landing trajectory. To maximize landing success,  \(K_p=0.5\) and \(K_d=0.1\) in \(c)\) proves to be the most suitable choice in this case, providing a good trade-off between the size of the landing success envelope and its uniformity.

\subsection{Wind evaluation results}

\begin{figure}[t]
    \centering
    \includegraphics[width=1\textwidth]{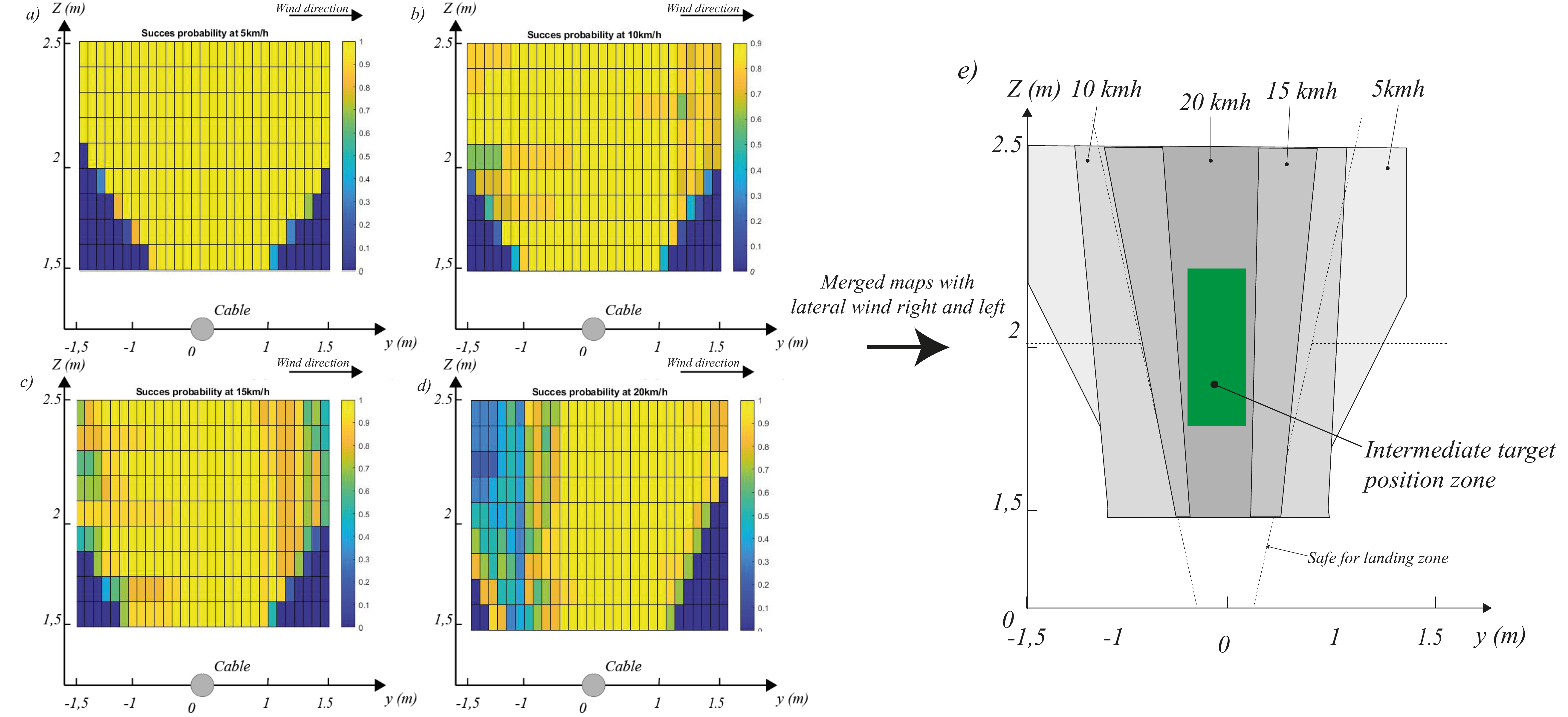}
    \caption{Maps of success in lateral wind from a) 5km/h, b) 10km/h, c) 15km/h and d) 20km/h can be merged in e) and can define the "intermediate target position zone" for alignment before landing}
    \label{fig:succes_map}
\end{figure}

The Monte Carlo simulations provide a detailed perspective on the landing envelope for the direct landing strategy (DLS) under different wind speeds, highlighting the initial position zones with the highest rates of successful landing initiation. Figure \ref{fig:succes_map} shows the map of landing success rates as a function of crosswind speed. The pilots in this study identify the worst-case scenario for landing when the crosswind direction goes from left to right on the maps, with success rates being symmetrical when the crosswind direction is reversed. The 100\% success rate area decreases with increasing wind speed. This area is primarily located above the power line and remains so up to wind speeds of at least 20 km/h. The zones with a 100\% success rate, considering the two worst wind directions together, are illustrated in Figure \ref{fig:succes_map}. 
The green area depicted in e) measuring 0.2m in width and 0.4m in height, represents the area with the highest landing success rate, and thus is the most preferable location to set the intermediate landing point. This area is defined based on the success rate and constraints associated with an overlapped cable environment under the worst-case scenario.

The results of these Monte Carlo simulations lay the groundwork for further refinement of the landing strategy, taking into account real-world variables such as environmental fluctuations, the geometry of the power lines, and drone performance. The implementation of this semi-autonomous landing strategy with a target before landing is projected to greatly improve the success rate and safety of drone operations during power line inspections.

\subsection{Strategies comparison}
In this subsection, we provide a comparative analysis of two distinct landing strategies: the Direct Landing Strategy (DLS) and the Two-Stage Landing Strategy (TSLS). These strategies differ in terms of their approach to aligning and landing the drone on high-voltage cables.

As a reminder, the DLS is a single-phase approach that directly descends the drone onto the cable, offering simplicity and time-efficiency but posing higher risks in terms of landing success and safety. On the other hand, the Two-Stage Landing Strategy (TSLS) employs a two-phase approach, consisting of an alignment phase and a landing phase. This enforces a specific approach trajectory, which in turn increases the confidence level of a successful landing.

\begin{figure}[H]
    \centering
    \includegraphics[width=0.75\textwidth]{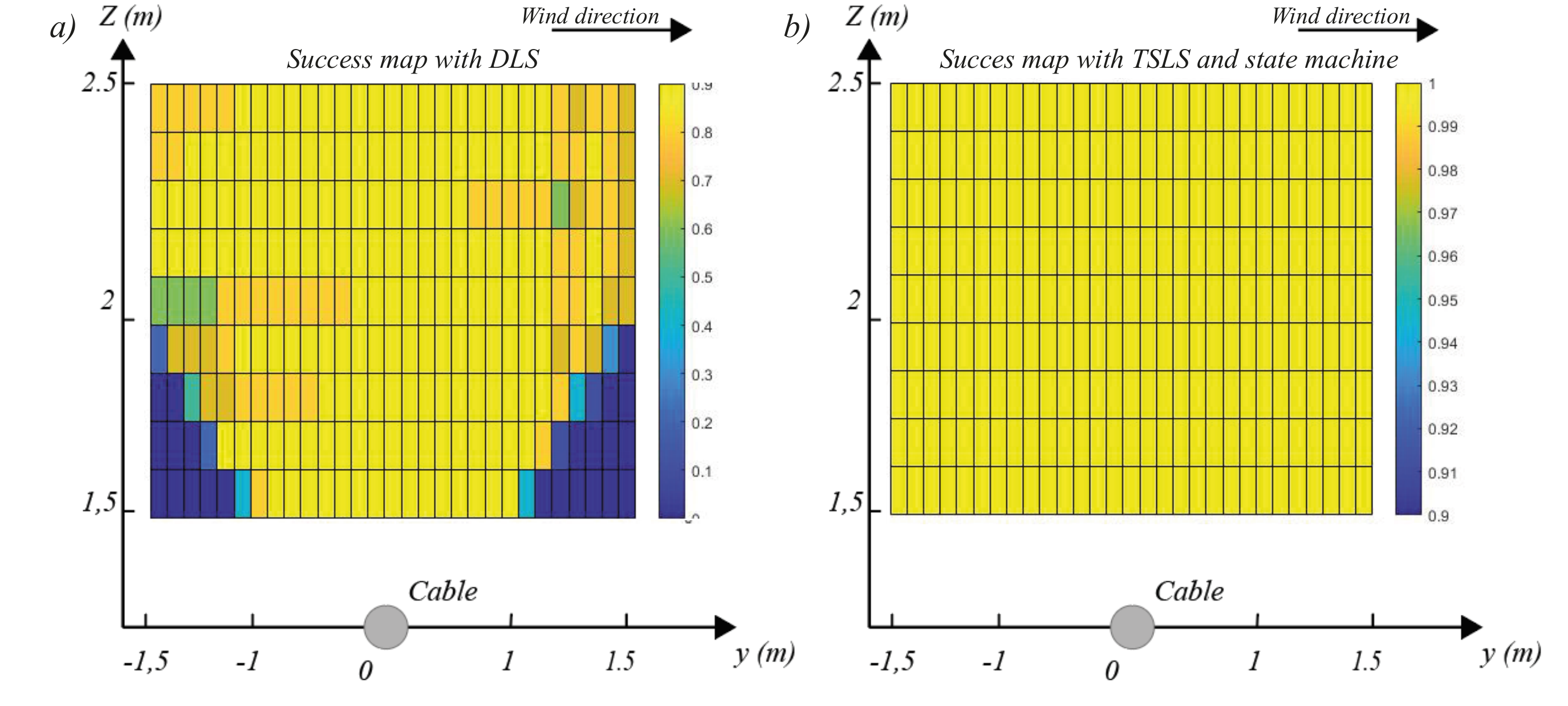}
    \caption{Success maps of landing strategies at 10km/h of lateral wind speed a) Direct landing strategy and b) Two-stage landing strategy and states machine. }
    \label{fig:MapsStrat}
\end{figure}

The comparison maps illustrated in Figure \ref{fig:MapsStrat} reveal a significant difference between the landing success rates of the two strategies. The TSLS map showcases a 100\% landing success rate for all initial positions above the cable with a 2m alignment target above the cable, emphasizing its effectiveness across various starting situations, more specifically in lower area where DLS have a zero success rate. Conversely, the DLS map indicates a lower success rate, stemming from its direct approach, which can result in potential risks and diminished accuracy.

In conclusion, the TSLS provides a considerably greater degree of landing success and safety compared to the DLS. By implementing the automatic TSLS with an alignment target position set at 2m above the cable in practical applications, operators can achieve more dependable and secure landings. This ultimately contributes to increased efficiency and safety in high-voltage cable inspections.

    \section{Field demonstration} \label{sec: fieldexp}
In this section, we present the results of the experimental testing of the landing algorithm on a set of real size un-energized power lines mock-up.


\subsection{Experimental setup}
\begin{figure}[H]
    \centering
    \includegraphics[width = 0.8\textwidth]{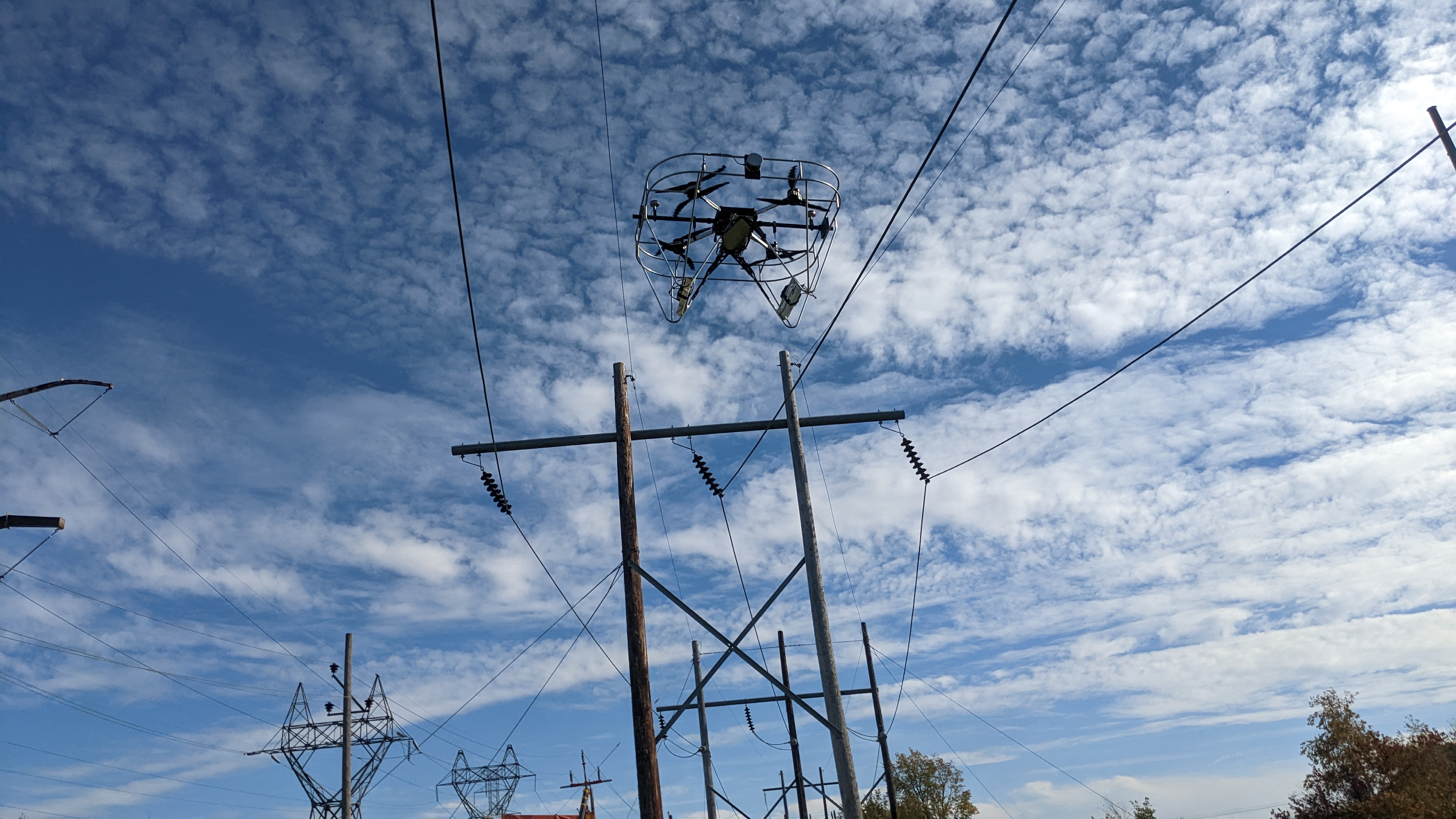}
    \caption{NADILE with a set of un-energized power lines mock-up.}
    \label{fig:Nadile_cables}
\end{figure}

The field tests were conducted on a set of un-energized power lines mock-up that replicate the conditions of actual high-voltage electrical cables, as show in Figure \ref{fig:Nadile_cables}. The drone was equipped with onboard sensors to record its states (position, orientation, velocity, etc.) during the landing attempts. Wind speed and direction were logged manually in a journal.

The NADILE was manually navigated to the "ready to align" zone above the target cable. Once in position, the landing procedure was initiated, and the drone's states were logged throughout the process. The landing approach was tested in various wind conditions, ranging from light winds (3-5 km/h) to more challenging windy conditions (20km/h with 25-30km/h wind gusts).

\subsection{Experimental results}
NADILE drone effectively aligning itself and landing securely on the un-energized power lines mock-up at 100\% in 50 attend. The performance of the controllers was consistent across various wind conditions, showcasing the robustness of the landing algorithm. The drone maintained stable flight during the alignment phase and demonstrated precise control over its position and orientation.

\begin{figure}[H]
    \centering
    \includegraphics[width=0.7\textwidth]{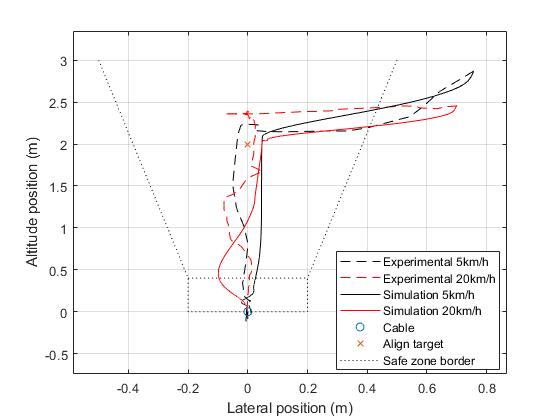}
    \caption{Experimental and simulation drone path comparison in y-z plane at 5km/h and 20km/h wind speed.}
    \label{fig:traj}
\end{figure}

The Two-Stage Landing Strategy (TSLS) behaves as anticipated in simulations, with a maximum error of 0.08 m even in challenging crosswind conditions of up to 20 km/h. This demonstrates the effectiveness of the algorithm in real-world scenarios. Figure \ref{fig:traj} illustrates both experimental and simulation landing trajectories from the 'ready to align' zone to the power line, in the presence of 5km/h and 20km/h lateral winds.

\begin{figure}[H]
    \centering
    \includegraphics[width=0.8\textwidth]{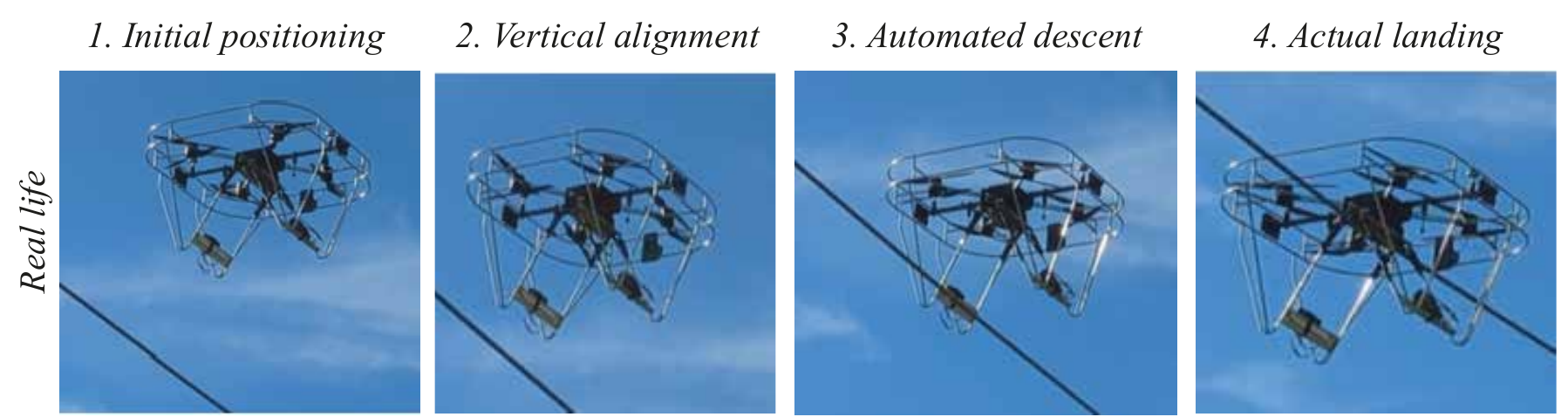}
    \caption{Landing sequence on an un-energized power lines mockup}
    \label{fig:land sequence}
\end{figure}

The Figure \ref{fig:land sequence} shows the landing sequence tested in simulation and outdoors on an experimental line under different wind conditions at four moments during the landing phase:
\begin{enumerate}
    \item Initial positioning of the drone on the side of the cable;
    \item Vertical alignment of the drone with the cable;
    \item Automated descent to the cable;
    \item Actual landing of the drone on the cable
\end{enumerate}

Field tests carried out on mock-ups of un-energized power lines offered insights into the NADILE's performance, including its wind resistance and cable alignment capabilities. These successful outcomes confirmed the potential of utilizing the drone for semi-autonomous landing on high-voltage cables in actual operational conditions. Future development and tests will persist in fine-tuning the algorithm and enhancing the drone's functionalities, with the goal of producing an increasingly proficient and dependable instrument for power line inspection and upkeep.
    \section{Conclusion}\label{sec:CCL}

In conclusion, this paper has addressed the challenges of manually landing a drone on power lines and introduced an experimental drone based on original LineDrone key features named NADILE. We have described two semi-autonomous landing strategies designed to facilitate the process of landing the drone on a power line, as well as the associated control schemes. Through the implementation of a dynamic drone simulation model resembling the actual NADILE and a Monte Carlo-based evaluation method, we were able to assess the success rates of initiating landing from different positions above the cable.

The results of these evaluations, including a comparison between the two landing strategies, have been presented, followed by a discussion on the outcomes. Real-world demonstrations of the concept were also shown to validate the findings. As demonstrated in this study, semi-autonomous landing in wind conditions is not only possible, but it also provides valuable assistance to drone pilots. Thanks to this system, pilots are equipped with the necessary tools to perform safe landings on power cables without the need to be positioned directly underneath the power lines. Therefore, the development of this semi-autonomous tool is essential for the successful deployment of the technology and to ensure the safety of pilots during inspections.

Looking ahead, further development of this system would ideally focus on creating a system that can withstand stronger winds while maintaining good control response. Additionally, it would be beneficial to develop a fully autonomous landing strategy, integrated with an autonomous navigation system, to further simplify the task of high-voltage power line inspection. This would potentially open the door to broader adoption and more effective use of drones in this critical application area.
    \subsection{Acknowledgments}We extend our sincere appreciation to the engineers and project managers at Hydro-Quebec and DroneVolt for their invaluable guidance and insights. Lastly, special recognition is due to our fellow students for their meaningful contributions that greatly enriched the development of this project.

\subsection{Competing interests statement}The authors declare there are no competing interests.

\subsection{Data availability}
The authors state that there is no raw data available due to data confidentiality.

\subsection{Author contributions}

Étienne Gendron was primarily responsible for the conceptualization, software and hardware development, validation, and original draft preparation. Marc-Antoine Leclerc contributed to the writing, review and editing, as well as developing the drone model and implementing the simulation for formal analysis. Samuel Hovington played a key role in software development and field testing, while also reviewing and editing the manuscript. Étienne Perron carried out formal analysis and investigation of the drone model, with David Rancourt providing necessary resources for the drone model study. Philippe Hamelin participated in the conceptualization, resource provision, manuscript review and editing, as well as project supervision. Alexis Lussier Desbiens, alongside Alexandre Girard, took charge of project conceptualization, methodology definition, and supervision, with Alexis further overseeing project administration and funding acquisition.


\subsection{Funding}This research was funded by Alliance grant number 2601-2600-703 and CRIAQ between University of Sherbrooke, Hydro-Québec, DroneVolt.

\subsection{Data availability statement}
Due to the commercial nature of this project and the associated proprietary restrictions, the raw data supporting the conclusions of this article will not be made available publicly. However, the authors confirm that the presented information and results are reproducible within the given constraints of the commercial agreement. Any inquiries regarding further specifics of the data and methodology should be directed towards the corresponding author, who will strive to address these to the best extent possible without violating any confidentiality agreements.
    \newpage
    \printbibliography
\end{document}